\documentclass[a4paper,10pt]{article}

\title{Modelling collisions in a relativistic plasma}
\author{Adam Noble$^{a,b}$ and David Burton$^{a,c}$ \\
\\
$^a$ Physics Department, Lancaster University, LA1 4YB, UK\\
$^b$ Department of Physics, SUPA, University of Strathclyde, G4 0NG, UK\\
$^c$ The Cockcroft Institute, Daresbury, UK}

\usepackage{amsmath,amssymb,bm}
\usepackage{cite,epsfig,color}
\newcommand{\xd}{\dot{x}}

\begin{document}

\maketitle

\begin{abstract}
Generalising the work of Lenard and Bernstein, we introduce a new, fully relativistic model to describe collisional plasmas.  Like the Fokker-Planck operator, this equation represents velocity diffusion and conserves particle number.  However, unlike the Fokker-Planck operator it is linear in the distribution function, and so more amenable to a fluid treatment.  By taking moments, we derive a new fluid model, and demonstrate the damping effects of collisions on Langmuir waves.

\end{abstract}




\section{Introduction}

In the high energy regime in which relativistic effects dominate -- such as laboratory based laser-plasma acceleration \cite{LWFA} -- plasmas are commonly described by the collisionless Vlasov equation.  This approximation is often justified as the timescales governing relativistic processes in an underdense plasma are typically much shorter than the average time between  collisions.  However, recent advances in high energy density science have increased the demand for descriptions of plasma dynamics fully incorporating both relativistic and collisional effects \cite{sim,Yu,sim2}.

Recent studies of laser-solid interactions \cite{experimental} and fast ignition fusion \cite{Weibel,CFI} sought to circumvent this issue by describing a relativistic electron beam interacting with a nonrelativistic collisional plasma.  However, this does not disguise the need for a fully consistent treatment of relativistic collisional plasmas.

The standard approach to describing plasma collisions is the Fokker-Planck equation \cite{RMJ,diff}, which supplements the Vlasov equation with a term representing diffusion in velocity space.  This nonlinear integral operator makes the Fokker-Planck equation cumbersome to work with in most cases of interest, and does not lend itself to the generation of fluid models.  As such, it is often replaced by simpler model collision operators \cite{Gross,Krook}.

The approach taken by Lenard and Bernstein \cite{LenBern} was to replace the drift and diffusion coefficients in the Fokker-Planck equation with simple, distribution independent forms.  This gave rise to a collision operator retaining three crucial features of the full Fokker-Planck description: it represents velocity diffusion; it conserves particle number; and it annihilates a Maxwellian equilibrium distribution.  For these reasons, it has been widely used in the literature \cite{Auerbach,Brodin-3,Ng}

Given the success of the Lenard-Bernstein equation, together with the current interest in relativistic collisional plasmas, it is somewhat surprising that a relativistic generalisation has not previously been considered.  In this paper, we provide such a generalisation, and explore some of its consequences.

In section~2 we introduce a relativistic collision operator, making full use of the geometry of the unit hyperboloid bundle, and demonstrate that it reduces to that of Lenard and Bernstein in the nonrelativistic limit. In section~3 we use this operator to derive a fluid model for a relativistic collisional plasma, and show how it modifies the dispersion relation for Langmuir waves.

Throughout, Latin indices run from $0$ to $3$, Greek indices from $1$ to $3$, and repeated indices are to be summed over their range.


\section{Linear relativistic collision operator}

The Vlasov equation for the electron distribution function, with additional terms representing collisions, may be written in the nonrelativistic limit as
\begin{equation}
 \frac{\partial f}{\partial t}+ v^\mu \frac{\partial f}{\partial x^\mu} + \frac{q}{m} (\bm{E} + \bm{v}\times \bm{B})^\mu \frac{\partial f}{\partial v^\mu}= \frac{\partial}{\partial v^\mu} \bigg( \bm{F}^\mu f + \bm{D}^{\mu \nu} \frac{\partial f}{\partial v^\nu} \bigg).  \label{FoPl}
\end{equation}
Here, $q<0$ and $m$ are respectively the charge and (rest) mass of the electron; $\bm{E}$ is the electric and $\bm{B}$ the magnetic field.
In the full Fokker-Planck treatment, the drift ($\bm{F}^\mu$) and diffusion ($\bm{D}^{\mu \nu}$) coefficients involve integrals over the 1-particle distribution $f$.  

Seeking a more tractable equation, Lenard and Bernstein \cite{LenBern} replaced the drift and diffusion coefficients in (\ref{FoPl}) with the simple distribution-independent forms
\begin{align}
\bm{F}^\mu  \rightarrow & \ \beta v^\mu \label{drift} \\
\bm{D}^{\mu \nu} \rightarrow&\  \alpha\beta \delta^{\mu\nu},  \label{diff}
\end{align}
where $\beta$ is an effective collision frequency and $\alpha$ is a measure of the equilibrium thermal velocity spread.

These operators retain three important features of the Fokker-Planck operator: they represent diffusion in velocity space; they conserve particle number,
\begin{equation}
\frac{\partial n}{\partial t} + \nabla \cdot (n {\bm u})=0,
\end{equation}
where $n$ is the zeroth and $n\bm u$ the first velocity moment of the distribution $f$; and they possess a Maxwellian equilibrium distribution,
\begin{equation}
f_M =A \exp(-{\bm v}^2/2\alpha).
\end{equation}

The Lenard-Bernstein equation, being much simpler than the full Fokker-Planck equation, has been used widely \cite{Auerbach,Brodin-3,Ng}.  Furthermore, as it is linear\footnote{That is, it is linear in $f$; however, it forms a nonlinear system when coupled to Maxwell's equations.}, its velocity moments provide a well-defined fluid model.  Given the current interest in relativistic phenomena in collisional plasmas \cite{sim,Yu,sim2,experimental,Weibel,CFI}, it is natural to seek a relativistic generalisation of (\ref{drift}, \ref{diff}).

The relativistic form of the LHS of (\ref{FoPl}) is commonly used in the literature on relativistic collisionless plasmas.  It has the form
\begin{equation}
Lf= \dot{x}^a \bigg( \frac{\partial f}{\partial x^a}- \frac{q}{m}F^\mu{}_a \frac{\partial f}{\partial v^\mu} \bigg),  \label{Vlasov}
\end{equation}
where $\xd^0=\sqrt{1+\bm v^2}$ and $\xd^\mu=v^\mu$ are the 4-velocity coordinates, $\bm v^2= \delta_{\mu \nu} v^\mu v^\nu$, and $F^\mu{}_a= \eta^{\mu b}F_{ba}$ are components of the electromagnetic 2-form.  $\eta^{ab}$ are components of the spacetime metric tensor, with signature $(-+++)$.

Using (\ref{diff}) in (\ref{FoPl}), the diffusion contribution to the collision operator is proportional to the Laplace-Beltrami operator on the flat space of nonrelativistic velocities.  The natural generalisation of this to the relativistic case is the Laplace-Beltrami operator $\Delta$ on the unit hyperboloid,  
\begin{equation}
\Delta f = 3 v^\mu \frac{\partial f}{\partial v^\mu}+ \bigg(\delta^{\mu \nu} + v^\mu v^\nu \bigg) \frac{\partial^2 f}{\partial v^\mu \partial v^\nu}.  \label{lap}
\end{equation}
See the appendix for a derivation of this expression.

It is less obvious how to generalise the drift term (\ref{drift}).  However, the closest relativistic operator which combines with (\ref{lap}) to annihilate the J\"uttner distribution 
\begin{equation}
f_J= A^\prime \exp( -\alpha^{-1} \sqrt{1+\bm v^2}),  \label{Jutt}
\end{equation} 
(the relativistic generalisation of the Maxwellian) is
\begin{equation}
Df= \sqrt{1+\bm v^2}\frac{\partial}{\partial v^\mu} \big( fv^\mu\big).  \label{D1}
\end{equation}

Hence a candidate for the relativistic generalisation of the Lenard-Bernstein equation (\ref{FoPl}--\ref{diff}) is
\begin{equation}
 \dot{x}^a \bigg( \frac{\partial f}{\partial x^a}- \frac{q}{m}F^\mu{}_a \frac{\partial f}{\partial v^\mu} \bigg)= \beta \Big( \sqrt{1+\bm v^2}\frac{\partial}{\partial v^\mu} \big( fv^\mu\big)+ \alpha \Delta f \Big).  \label{rel}
\end{equation}

\subsection{Relativistic covariance}

Although (\ref{rel}) is relativistic, it is not manifestly covariant.  The drift term is not invariant under Lorentz transformations, and the diffusion term, being the Laplace-Beltrami operator on the unit hyperboloid (not the unit hyperboloid bundle) is also not guaranteed to be form-invariant.  Only by recasting (\ref{rel}) as a fully covariant equation can we be sure it is well-defined.

To make (\ref{Vlasov}) covariant, the Liouville vector $L$ can be written in terms of horizontal and vertical lifts of structures on spacetime $\cal M$ to the tangent bundle $T\cal M$ (see \cite{Yano} for details):
\begin{equation}
 L= \dot{x}^a \bigg[ \Big(\frac{\partial}{\partial x^a}\Big)^{\bm{H}}- \Big(\frac{q}{m}F^b{}_a \frac{\partial}{\partial x^b}\Big)^{\bm{V}} \bigg].  \label{Liouville}
\end{equation}

It is important to recognise that, although (\ref{Liouville}) is defined on the whole of the tangent bundle, we can nevertheless restrict attention to the physically meaningful unit hyperboloid bundle
\begin{equation}
 {\cal H}= \{ (x,\xd) \in T{\cal M} : \varphi(x,\xd)=0 \}, \qquad \varphi \equiv \eta^{\bm V}_{ab}(x) \xd^a \xd^b +1.
\end{equation}
This follows since the Liouville vector field is tangent to $\cal H$, $L\varphi=0$.

That (\ref{D1}) is not covariant is to be expected, since by construction the collision operator annihilates the J\"uttner distribution (\ref{Jutt}), which itself is not covariant.  However, the J\"uttner distribution can be written covariantly on the tangent bundle  
\begin{equation}
f= A^\prime \exp( -\alpha^{-1} Y_a \dot{x}^a),
\end{equation} 
at the expense of introducing the 1-form $Y= Y_a dx^a=dx^0$.

Expanding the derivative in (\ref{D1}), we can write
\begin{equation}
Df= 3\beta \sqrt{1+\bm v^2} f+ \beta \sqrt{1+\bm v^2} v^\mu \frac{\partial f}{\partial v^\mu}.
\end{equation}
Both the scalar $\sqrt{1+\bm v^2}$ and the vector $\sqrt{1+\bm v^2}v^\mu \partial/\partial v^\mu$ can be expressed in terms of the 1-form $Y$, and maps from spacetime to the unit hyperboloid.

Let $\Sigma_x : {\cal H}_x \rightarrow T{\cal M}$ be the embedding map of the unit hyperboloid over $x \in \cal M$ into the tangent bundle, and define the map
\begin{eqnarray}
\nonumber \iota : \Lambda^1 {\cal M} &\rightarrow& \Lambda^0 T{\cal M} \\
\omega_a dx^a & \mapsto & \omega^{\bm V}_a \dot{x}^a.
\end{eqnarray}
Then
\begin{align}
\sqrt{1+\bm v^2} =& \Sigma^\ast_x \iota Y, \\
\sqrt{1+\bm v^2} v^\mu \frac{\partial }{\partial v^\mu}=& (d\Sigma^\ast_x \iota Y)^\sharp,
\end{align}
where $\Sigma^\ast_x$ is the pull-back by $\Sigma_x$, and $\sharp$ is the dual with respect to the metric $g$ on ${\cal H}_x$, that is, $g(\omega^\sharp, X)= \omega(X)$ for all vectors $X$ on ${\cal H}_x$.


Then, defining $f_x = \Sigma^\ast_x f$, the relativistic Lenard-Bernstein equation (\ref{rel}) can be written
\begin{equation}
 \Sigma^\ast_x Lf= \beta \Big(3\Sigma^\ast_x \iota Y  + (d\Sigma^\ast_x \iota Y)^\sharp + \alpha\Delta \Big) f_x.   \label{cov}
\end{equation}
Since each term in this equation is defined intrinsically, it follows that (\ref{cov}) is fully covariant, and hence (\ref{rel}) is well-defined.

It remains to interpret the 1-form $Y$.  This represents the frame in which the equilibrium J\"uttner distribution is `at rest' (that is, its first velocity moment has no spatial component).  Given the role of the background ions in determining this frame, it is natural to identify $Y$ with the velocity field of the ions, $Y= -\widetilde{V}_{\text{ion}}$.  The metric dual of a vector $X$ on $\cal M$ is defined by $\widetilde{X}(Z)=\eta(X,Z)$ for all vectors $Z$ on {$\cal M$}, $\eta$ being the spacetime metric.

\subsection{Nonrelativistic limit}

Having obtained a linear equation for the 1-particle distribution incorporating collisions, we should confirm that it does indeed reproduce (\ref{FoPl}--\ref{diff}) in the nonrelativistic limit.  To do this, we introduce a small parameter $\varepsilon$, and define the scaled velocities
\begin{equation}
 v^\mu = \varepsilon \hat{v}^\mu.   \label{scalev}
\end{equation}

Taking the typical time scale to be of order unity, (\ref{scalev}) implies that typical spatial scales should scale with $\varepsilon$, so introduce rescaled coordinates
\begin{equation}
 x^0=\hat{x}^0, \quad x^\mu =\varepsilon \hat{x}^\mu.
\end{equation}

In the nonrelativistic limit, we expect $\alpha$ to correspond to the square of the thermal velocity spread, and $\beta$ to the collision frequency, which should be the inverse of the typical time scale.  Hence
\begin{equation}
 \alpha = \varepsilon^2 \hat{\alpha}, \qquad \beta= \hat{\beta}.  
\end{equation}

Finally, if the electric and magnetic contributions to the Lorentz force are to be comparable, components of the electromagnetic 2-form should scale as
\begin{equation}
 F^\mu{}_\nu = \hat{F}^\mu{}_\nu, \quad F^0{}_\mu= \varepsilon \hat{F}^0{}_\mu.
\end{equation}

Then taking the limit $\varepsilon \rightarrow 0$, keeping circumflexed quantities constant, (\ref{rel}) becomes
\begin{equation}
 \frac{\partial f}{\partial \hat{x}^0} + \hat{v}^\mu \frac{\partial f}{\partial \hat{x}^\mu}- \frac{q}{m} \big(\hat{F}^\mu{}_0 + \hat{F}^\mu{}_\nu \hat{v}^\nu\big) \frac{\partial f}{\partial \hat{v}^\mu}= \hat{\beta} \frac{\partial}{\partial \hat{v}^\mu} \Big( f\hat{v}^\mu+ \hat{\alpha} \delta^{\mu \nu} \frac{\partial f}{\partial \hat{v}^\nu}\Big),
\end{equation}
which is indeed (\ref{FoPl}--\ref{diff}).

\section{Macroscopic fluid models}

Equation (\ref{rel}) forms a closed system when coupled to Maxwell's equations, using the electric current components
\begin{equation}
 j^a = q \int f \xd^a \frac{d^3v}{\sqrt{1+\bm v^2}}- qN^a_\text{ion},
\end{equation}
where the ion number current $N^a_\text{ion}=-n_\text{ion} \eta^{ab}Y_b$, with $n_\text{ion}$ the (constant) ion number density.  However, it is not always convenient to work with differential equations on the 7-dimensional unit hyperboloid bundle, and solving for the full distribution function is often unnecessary.

Often it can both be more convenient and facilitate interpretation to work with a fluid model on spacetime rather than a kinetic model on the unit hyperboloid bundle.  Such a fluid model may be constructed by taking velocity moments of the kinetic equation.  Unlike the full Fokker-Planck equation, the relativistic Lenard-Bernstein equation (\ref{rel}) is naturally suited to such a construction.

With the usual definition of the number current and the stress-energy tensor as respectively the first and second moments of the particle distribution,
\begin{equation}
 n^a= \int f\xd^a \frac{d^3 v}{\sqrt{1+{\bm v}^2}}, \qquad S^{ab}= m\int f\xd^a \xd^b \frac{d^3 v}{\sqrt{1+{\bm v}^2}},
\end{equation}
the integral of the relativistic Lenard-Bernstein equation (\ref{rel}) shows that particle number is conserved:\footnote{This equation is valid in an inertial coordinate system.  In an arbitrary coordinate system $\nabla_a n^a=0$, with $\nabla_a$ the covariant derivative.  Similar comments apply to the other fluid equations.}
\begin{equation}
 \partial_a n^a=0.   \label{cons}
\end{equation}

That the drift term does not contribute a source to the RHS of (\ref{cons}) is evident, as it is the integral of a divergence, and $f$ is assumed to vanish at infinity.  The same can be seen of the diffusion term, when it is recognised that the Laplace-Beltrami operator can be written
\begin{equation}
 \Delta f= \frac{1}{\sqrt{\text{det}\ g}} \frac{\partial}{\partial v^\mu} \Big( \sqrt{\text{det} \ g} \ g^{\mu \nu} \frac{\partial f}{\partial v^\nu} \Big),
\end{equation}
where $g$ is the metric on ${\cal H}_x$ (see appendix), and the determinant of this metric is
\begin{equation}
 \text{det}\ g= \frac{1}{1+ \bm v^2}.
\end{equation}

Multiplying (\ref{rel}) by $m\xd^b$ before integrating yields
\begin{equation}
 \partial_a S^{ab}= -qF^b{}_a n^a + \big(\partial_a S^{ab}\big)_{\text{drift}} +\big(\partial_a S^{ab}\big)_{\text{diff}}.
\end{equation}

The diffusion contribution to the divergence of the stress-tensor is readily calculated using Green's second identity:
\begin{align}
 \nonumber \big(\partial_a S^{ab}\big)_{\text{diff}}&\equiv m\alpha \beta \int \xd^b \Delta f \frac{d^3 v}{\sqrt{1+\bm v^2}}\\
\nonumber &= m\alpha \beta \int \Delta \xd^b f \frac{d^3 v}{\sqrt{1+\bm v^2}}\\
&= 3m\alpha \beta n^b,
\end{align}
where the result $\Delta \xd^a= 3\xd^a$ has been used (see appendix for details).

The contribution of the drift term to the divergence of the stress-energy tensor can be integrated by parts (again neglecting boundary terms by assuming $f$ and its derivatives fall off sufficiently rapidly for large $v$):
\begin{align}
\nonumber \big(\partial_a S^{ab}\big)_{\text{drift}}&\equiv m\beta \int \xd^b \sqrt{1+\bm v^2} \frac{\partial}{\partial v^\mu}(fv^\mu) \frac{d^3 v}{\sqrt{1+\bm v^2}}\\
&= \beta\big( \eta^{ac}Y^b-\eta^{bc}Y^a\big) S_{ac},
\end{align}
where $Y^a= \eta^{ab} Y_b$ are components of the 1-form $Y=dx^0$ representing the velocity of the background ions.

The full equation for the divergence of the stress-energy tensor then is
\begin{equation}
 \partial_a S^{ab}= -qF^b{}_a n^a +\beta\big( \eta^{ac}Y^b-\eta^{bc}Y^a\big) S_{ac}+ 3m\alpha \beta n^b.  \label{div}
\end{equation}
Note that the term involving the equilibrium temperature $\alpha$, which originates in the curvature of ${\cal H}_x$, is a purely relativistic effect, vanishing in the nonrelativistic limit.

In practice, it is useful to recast the fluid equations in terms of the average velocity field
\begin{equation}
 u^a= h^{-1} n^a,  \label{defu}
\end{equation}
where $h$ is the zeroth moment of the distribution
\begin{equation}
 h= \int f \frac{d^3 v}{\sqrt{1+ \bm v^2}},
\end{equation}
and the pressure tensor
\begin{equation}
 P^{ab}=m\int f(\dot{x}^a-u^a) (\dot{x}^b-u^b) \frac{d^3 v}{\sqrt{1+\bm v^2}}.
\end{equation}
Then (\ref{div}) becomes
\begin{equation}
 u^b \partial_b u^a= -\frac{1}{m h}\partial_b P^{ab} -\frac{q}{m}F^a{}_b u^b 
-\beta\big( \delta^a_b+ u^au_b+ \frac{1}{mh}P^a{}_b \big) Y^b + 3\alpha \beta u^a.   \label{force}
\end{equation}

Note that the average velocity $u^a$ is {\em not} unit normalised, but rather its norm is related to the trace of the pressure tensor:
\begin{equation}
 \eta_{ab}u^a u^b= -(1+ \frac{1}{mh}P^a{}_a).  \label{norm}
\end{equation}

In common with the fluid equations derived from the collisionless Vlasov equation, (\ref{cons}, \ref{div}) (or equivalently, (\ref{cons}, \ref{defu}, \ref{force})) do not form a closed system, but must be supplemented by an equation of state relating $S^{ab}$ and $n^a$ (or relating $P^{ab}$, $h$ and $u^a$).  

\subsection{Langmuir waves in a cold plasma}

One commonly used equation of state is the vanishing of the pressure tensor, representing a cold plasma.  Since $\alpha$ is a measure of the equilibrium thermal velocity spread, it should be zero for a cold plasma.  Indeed, setting $P^{ab}=0$ and contracting (\ref{force}) with $u_a$ leads to the consistency requirement $\alpha=0$.  As an illustration of the effects of collisions, consider Langmuir waves in a cold plasma.

Assuming a cold plasma, (\ref{force}, \ref{norm}) become
\begin{equation}
 \nabla_u \widetilde{u} = \frac{q}{m}i_u F- \beta \big( Y+ Y(u) \widetilde{u} \big), \qquad \eta(u,u)=-1,  \label{acc}
\end{equation}
where $u= u^a \partial/\partial x^a$, and $\nabla$ is the Levi-Civita connection on $\cal M$.  The metric dual of a vector $X$ is defined by $\widetilde{X}(Z)=\eta(X,Z)$ for all vectors $Z$ on $\cal M$.  (\ref{acc})  then couples to Maxwell's equations as
\begin{equation}
 d\star F= -q\star (n \widetilde{u}+ n_\text{ion} Y), \qquad dF=0,  \label{Max}
\end{equation}
with $n=\sqrt{-\eta_{ab}n^a n^b}$ the electron proper number density.  The contribution of the collisions to (\ref{acc}) is proportional the 3-velocity of the ions as measured in the rest frame of the electrons, and may be interpreted as a friction force.

The simplest solution to (\ref{acc}, \ref{Max}), representing electrons at rest and vanishing electromagnetic field, is
\begin{equation}
 n=n_\text{ion}, \qquad u= -\widetilde{Y}, \qquad F=0.  \label{equ}
\end{equation}
The metric dual of a 1-form $A$ on $\cal M$ is inverse to that of vectors: $\eta(\widetilde{A},Z)=A(Z)$ for all vectors $Z$ on $\cal M$.  

To see how collisions influence the plasma dynamics, perturb about (\ref{equ}),
\begin{equation}
 u=-\widetilde{Y}+ \varepsilon \Upsilon, \qquad n= n_\text{ion} + \varepsilon \nu, \qquad F= \varepsilon \Phi,
\end{equation}
and to leading order in $\varepsilon$ (\ref{acc}, \ref{Max}) become

  \begin{equation}
\nonumber  \nabla_{\widetilde{Y}} \Upsilon= \frac{q}{m} \widetilde{i_{\widetilde{Y}}\Phi}- \beta \Upsilon, \qquad Y(\Upsilon)=0,
 \end{equation}
\begin{equation}
 d\star\Phi= q\star \big( \nu Y- n_\text{ion} \widetilde{\Upsilon} \big), \qquad d\Phi=0.  \label{linear}
\end{equation}

Langmuir waves are solutions to (\ref{linear}) representing longitudinal oscillations, i.e. $\Phi= Edt\wedge dz$, $\Upsilon= \lambda \partial/\partial z$, with $E$, $\lambda$ and $\nu$ functions of $(t,z)$ only.  Assuming a temporal dependence $\exp(-i\omega t)$, (\ref{linear}) yields the dispersion relation
\begin{equation}
 \omega= -i\frac{\beta}{2} \pm \sqrt{\omega^2_p- \Big(\frac{\beta}{2}\Big)^2},   \label{disp}
\end{equation}
where $\omega_p= \sqrt{q^2n_\text{ion}/m}$ is the plasma frequency.  Thus, the effect of the collisions is to downshift the frequency of the oscillations and damp their amplitude.  The dispersion relation (\ref{disp}) has previously been derived for high velocity waves in \cite{Lat}, using the nonrelativistic Krook collision operator.

\section{Conclusion}
It is becoming increasingly necessary to describe the dynamics of a plasma in which both collisional and relativistic effects are important.  The full relativistic Fokker-Planck equation is for many purposes too cumbersome, so simpler models are sought.

We have presented one such model that retains many of the defining features of the Fokker-Planck equation, but which remains simple enough to yield useful information about collective motion.  This model may be regarded as a relativistic generalisation of the Lenard-Bernstein equation.

By taking velocity moments of this equation, we have generated a set of fluid equations, which when supplemented with equations of state provide a description of a relativistic collisional plasma in terms of fields on spacetime.  As an illustration of this equation, we have shown how collisions modify the dispersion relation for Langmuir waves in a cold plasma.

\section*{Acknowledgements}
We would like to thank Bernhard Ersfeld, Dino Jaroszynski and Raoul Trines for useful discussions.

\appendix
\section{Geometry of the unit hyperboloid}

The spacetime metric $\eta= \eta_{ab}dx^a \otimes dx^b$ can be lifted onto the tangent bundle $T{\cal M}$ to give the Sasaki metric \cite{Yano}
\begin{equation}
 \eta^{\bm D}= \eta_{ab} \Big( dx^{a\bm H} \otimes dx^{b\bm H} +dx^{a\bm V} \otimes dx^{b\bm V} \Big),
\end{equation}
where $\bm H$ and $\bm V$ are respectively the horizontal and vertical lifts \cite{GCM8}.  
Using the embedding map $\Sigma_x$ of the unit hyperboloid over $x \in \cal M$ into $T\cal M$, this gives rise to a metric on ${\cal H}_x$:
\begin{equation}
 g= \Sigma^\ast_x \eta^{\bm D}= \big( \delta_{\mu\nu} - \frac{v_\mu v_\nu}{1+\bm v^2 } \big) dv^\mu \otimes dv^\nu,
\end{equation}
where $v_\mu= \delta_{\mu \nu} v^\nu$.

The Laplace-Beltrami operator on the unit hyperboloid ${\cal H}_x$ acting on scalars can be written as
\begin{equation}
\Delta f= \# d\# df,
\end{equation}
where $\#$ is the Hodge map of $g$.  Expanding this equation, we have
\begin{equation}
\Delta f= \frac{\partial f}{\partial v^\mu} \# d\# dv^\mu + \frac{\partial^2 f}{\partial v^\mu \partial v^\nu} \# (dv^\mu \wedge \# dv^\nu).  \label{Lap}
\end{equation}

We can evaluate the terms involving the Hodge map using the orthonormal basis and volume 3-form
\begin{equation}
 e^1= \frac{d\rho}{\sqrt{1+\rho^2}}, \qquad e^2= \rho d\theta, \qquad e^3= \rho \sin\theta d\varphi,  \qquad \# 1= e^1 \wedge e^2 \wedge e^3,
\end{equation}
and the relations
\begin{equation}
 v^1= \rho \sin \theta \cos \varphi, \qquad v^2 = \rho \sin \theta \sin \varphi, \qquad v^3 =\rho\cos\theta.
\end{equation}

All the terms in (\ref{Lap}) involving the Hodge map can be calculated explicitly.  However, due to the isotropy of ${\cal H}_x$, it is sufficient to find $\# d\# dv^3$, $\#(dv^3 \wedge \# dv^3)$ and $\#(dv^1 \wedge \# dv^3)$.

We have:
\begin{align}
\nonumber dv^3&= \cos\theta d\rho - \rho\sin \theta d\theta\\
&= \sqrt{1+\rho^2} \cos\theta e^1 - \sin\theta e^2.  \label{dvz}\\
\nonumber \# dv^3 &= \sqrt{1+\rho^2} \cos\theta e^2 \wedge e^3 - \sin\theta e^3 \wedge e^1\\
&= \rho^2\sqrt{1+\rho^2} \sin\theta \cos\theta d\theta \wedge d\varphi - \frac{\rho}{\sqrt{1+\rho^2}}\sin^2\theta d\varphi \wedge d\rho.\\
\nonumber d\# dv^3&= 3\rho\cos\theta \frac{\rho^2}{\sqrt{1+\rho^2}} \sin\theta d\rho \wedge d\theta \wedge d\varphi \\
&= 3v^3 \# 1.
\end{align}

From (\ref{dvz}) and $e^\mu \wedge \# e^\nu= \delta^{\mu \nu} \#1$, it follows that
\begin{equation}
 \# (dv^3 \wedge \# dv^3)= 1+ \rho^2 \cos^2 \theta= 1+ (v^3)^2,
\end{equation}
and using
\begin{align}
 \nonumber dv^1 &= \sin\theta \cos \varphi d\rho + \rho \cos\theta \cos \varphi d\theta - \rho \sin\theta \sin\varphi d\varphi\\
&= \sqrt{1+\rho^2}\sin\theta \cos\varphi e^1+ \cos\theta \cos\varphi e^2 - \sin\varphi e^3,
\end{align}
we have
\begin{equation}
\# (dv^1 \wedge \# dv^3)= \rho^2 \sin\theta \cos\theta \cos\varphi= v^1 v^3.
\end{equation}
These results can be readily extended, by direct calculation or symmetry arguments, to obtain
\begin{align}
 &\# d\# dv^\mu= 3v^\mu,\\
 &\# (dv^\mu \wedge \# dv^\nu) = \delta^{\mu\nu} + v^\mu v^\nu.
\end{align}
Substituting these into (\ref{Lap}) yields (\ref{lap}).

\end{document}